\documentclass[twoside]{article}
\usepackage{kyber2e}
\usepackage{graphicx}
\usepackage{amsmath}
\usepackage{amsfonts}
\usepackage{amssymb}
\usepackage{cite}

\def \hs {\hspace*{-2mm}}
\newfont{\mi}{cmti9}
\newfont{\m}{cmr8}
\newfont{\ms}{cmsl8}
\newfont{\autor}{cmcsc10}
\newtheorem{theorem}{Theorem}

\newtheorem{lemma}[theorem]{Lemma}

\begin{document}

%povinna titulni strana - Kybernetika
\rule[3mm]{128mm}{0mm} \vspace*{-16mm}

{\footnotesize K\,Y\,B\,E\,R\,N\,E\,T\,I\,K\,A\, --- \,V\,O\,L\,U\,M\,E\, 
\textit{4\,1\,} (\,2\,0\,0\,9\,)\,,\, N\,U\,M\,B\,E\,R\, x\,,\,\
P\,A\,G\,E\,S\, \,x\,x\,x -- x\,x\,x}\newline
\rule[3mm]{128mm}{0.2mm}

\vspace*{11mm}

{\large \textbf{\noindent JOINT\, RANGE\, OF\, R\'{E}NYI\, ENTROPIES}}

\vspace*{8mm}

{\autor\indent Peter Harremo\"{e}s }

\vspace*{23mm}

{\small The exact range of the joined values of several R\'{e}nyi entropies
is determined. The method is based on topology with special emphasis on the
orientation of the objects studied. Like in the case when only two orders of
R\'{e}nyi entropies are studied one can parametrize upper and lower bounds
but an explicit formula for a tight upper or lower bound cannot be given.}

\smallskip

\begin{minipage}[t]{112mm}
R\'{e}nyi entropies, Shannon entropy, information diagram.
\end{minipage}\smallskip


\section{\protect\normalsize \hs INTRODUCTION}

{\normalsize Let $P=\left( p_{1},p_{2},...,p_{n}\right) $ be a probability
vectors. For $\alpha\in\mathbb{R}\backslash\{0,1\}$ the R\'{e}nyi entropy of 
$P$ of order $\alpha$ is defined as a number in $\left[ 0;\infty\right] $
given by the equation%
\begin{equation*}
H_{\alpha}\left( P\right) =\frac{1}{1-\alpha}\log\left(
\sum_{i}p_{i}^{\alpha}\right) .
\end{equation*}
This definition is extended by continuity so that 
\begin{align*}
H_{-\infty}\left( P\right) & =-\log\min_{i}p_{i}~; \\
H_{0}\left( P\right) & =\log(\text{number of }p_{i}\not =0); \\
H_{1}\left( P\right) & =-\sum_{i}p_{i}\log p_{i}~; \\
H_{\infty}\left( P\right) & =-\log\max_{i}p_{i}~.
\end{align*}
The R\'{e}nyi entropy $H_{0}$ is essentially the Hartley entropy, and was
one among other sources of inspiration to Shannon's information theory. The R%
\'{e}nyi entropy of order $\infty$ is also called the min-entropy and
essentially related to the "probability of error". The R\'{e}nyi entropy $%
H_{2}$ is related to index of coincidence and other quantities used for
special purposes in crypto analysis, physics etc. \cite{Harremoes2001h,
Arndt2001}. }

{\normalsize For all $\alpha$ the R\'{e}nyi entropy $H_{\alpha}$ has the
nice property of being additive on product measures. In noiseless source
coding for finite systems one wants to avoid very long code words. For such
systems the R\'{e}nyi entropy of some order $\alpha<1$ (depending on the
memory of the system) determines how much the source can be compressed. R%
\'{e}nyi entropies are also related to general cut-off rates and "guess-work
moments" \cite{Csiszar1995, Arikan1996}. }

{\normalsize The relation between $H_{0}$ and $H_{1}$ is given by the simple
inequality%
\begin{equation*}
H_{1}\left( P\right) \leq H_{0}\left( P\right) .
\end{equation*}
This is a special case of the general result that}%
\begin{equation*}
\alpha\rightarrow H_{\alpha}\left( P\right)
\end{equation*}
{\normalsize is a strictly decreasing function except for uniform
distributions where it is constant, which follows from a simple application
of Jensen's Inequality. The relation between $H_{1}$ and $H_{\infty}$ has
been determined independently in various articles \cite{Kovalevskij1967,
Tebbe1968, Ben-Bassat1978, Golic1987, Feder1994}. The relation between
Shannon entropy and $H_{2}$ has been studied in \cite{Gyorgy2000} and in
more detail in \cite{Harremoes2001h}. The result is illustrated on Figure \ref%
{Fig0} and by the following theorem. }

\begin{theorem}
{\normalsize The the upper bound on $H_{2}\left( P\right) $ given $%
H_{1}\left( P\right) $ is attained by a mixture of uniform distributions on $%
k$ and $k+1$ points where $k$ is determined by the condition $\log k\leq
H_{1}\left( P\right) <\log \left( k+1\right) .$ The lower bound on $%
H_{2}\left( P\right) $ is attained by a mixture of the uniform distribution
on $n$ points and a uniform distribution on a singleton.

\begin{figure}[htp]
\centering
\includegraphics[width=3in]{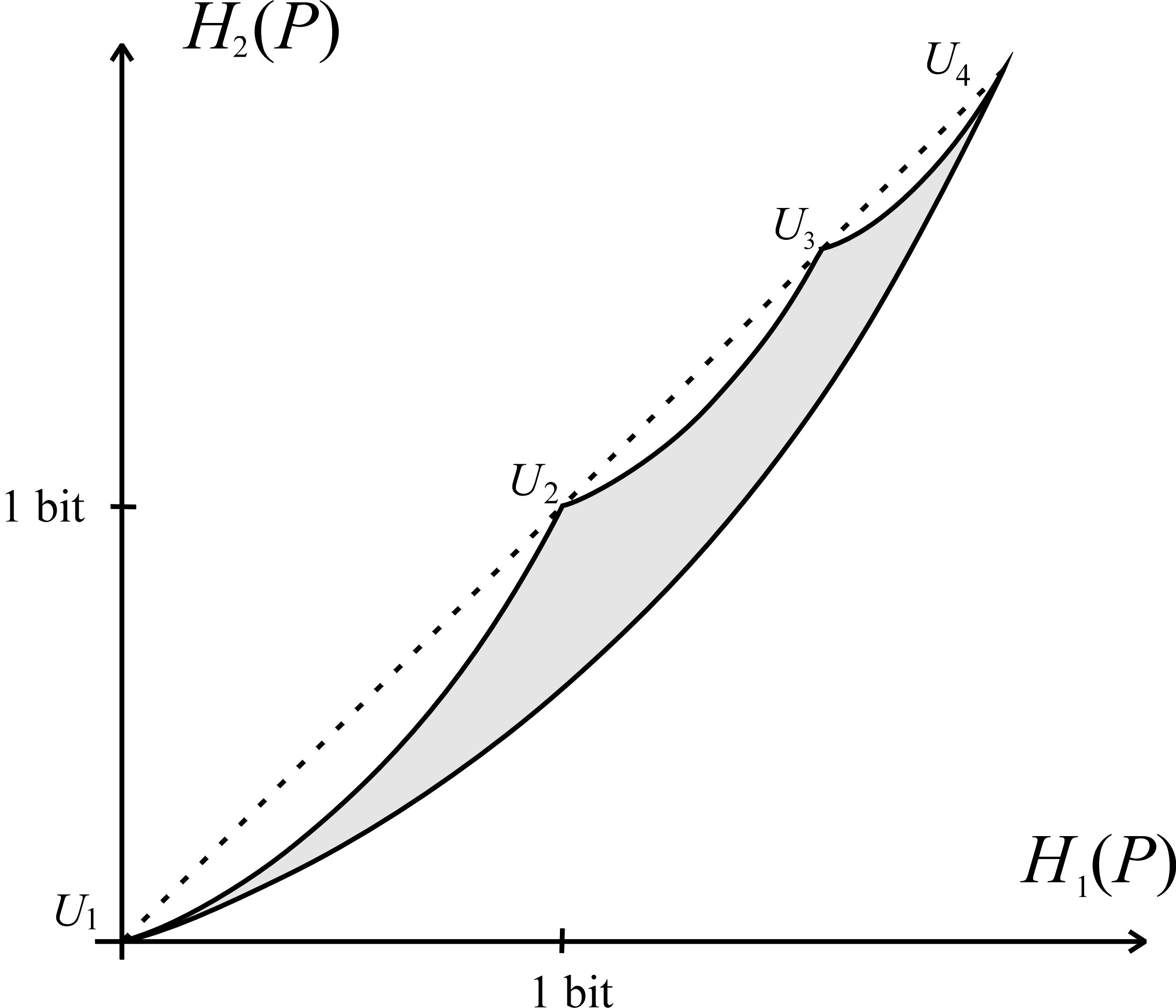}
\caption{Range of $P\rightarrow \left( H_{1}\left( P\right)
,H_{2}\left( P\right) \right) $ for a four element set.}\label{Fig0}
\end{figure}

%\FRAME{ftbpFU}{%
%2.9092in}{2.495in}{0pt}{\Qcb{Range $P\rightarrow \left( H_{1}\left( P\right)
%,H_{2}\left( P\right) \right) $ for a four element set.}}{}{joint.jpg}{%
%\special{language "Scientific Word";type "GRAPHIC";maintain-aspect-ratio
%TRUE;display "USEDEF";valid_file "F";width 2.9092in;height 2.495in;depth
%0pt;original-width 8.3662in;original-height 7.1736in;cropleft "0";croptop
%"1";cropright "1";cropbottom "0";filename 'joint.jpg';file-properties
%"XNPEU";}
%}
}
\end{theorem}

{\normalsize In this paper we shall generalize this result and determine the
joint range of several R\'{e}nyi entropies. In general the boundary can be
parametrized, but upper and lower bounds cannot be given by explicit
formulas. The reason is that the inverse of the function $s\rightarrow
H_{\alpha}\left( sU_{k}+\left( 1-s\right) U_{k+1}\right) ,$ where $U_{k}$
and $U_{k+1}$ are uniform distributions, is in general not an elementary
function. }

{\normalsize Recently the joint range of R\'{e}nyi entropies has been used
to determine the relative Bahadur efficiency of various power divergence
statistics \cite{Harremoes2008, Harremoes2008e}. In these papers the joint
range of $H_{1}$ and $H_{\alpha}$ was used with a reference to \cite%
{Harremoes2001h} where the general result for comparison of two R\'{e}nyi
entropies was mentioned without proof. In some cases in physics, joint
values of $H_{2}\left( P\right) $ and $H_{3}\left( P\right) $ can be
measured or computed and one is interested in bounds on $H_{1}$ \cite%
{Zyczkowski2003}. In order to get bounds on $H_{1}$ one is interested in the
exact range of the mapping 
\begin{equation*}
\Psi:P\rightarrow\left( H_{1}\left( P\right) ,H_{2}\left( P\right)
,H_{3}\left( P\right) \right) .
\end{equation*}
In this paper the methods developed in \cite{Harremoes2001h} will be refined
in order to be able to describe the joint range of in principle any number
of R\'{e}nyi entropies of positive order. We restrict our attention to
non-negative orders because these are the most important for applications
and because R\'{e}nyi entropies of negative orders are not continuous near
uniform distributions. Although the method is very general we shall only go
into details in the cases where two or three R\'{e}nyi entropies are
compared. The main result is that the range has a boundary that can be
parametrized by certain mixtures of uniform distributions. }

\section{\protect\normalsize \hs REDUCTION TO MIXTURES OF UNIFORM
DISTRIBUTIONS}

{\normalsize A probability vector $P$ on a set with }$n$ {\normalsize %
elements can be parametrized by its point probabilities as $\left(
p_{1},p_{2},...,p_{n}\right) $ where $p_{j}\geq0$ and }%
\begin{equation*}
{\sum_{j=1}^{n}p_{j}=1.}
\end{equation*}
{\normalsize Here we shall assume that $n$ is fixed so that that $%
H_{0}\left( P\right) \leq\log n.$ In order to study the range of $%
P\curvearrowright \left( H_{\alpha_{1}}\left( P\right) ,H_{\alpha_{2}}\left(
P\right) ,\cdots,H_{\alpha_{m}}\left( P\right) \right) $ we first consider
the related map%
\begin{equation}
P\rightarrow\left( 
\begin{array}{c}
\frac{1}{1-\alpha_{1}}\log\left( \sum p_{j}^{\alpha_{1}}\right) \\ 
\frac{1}{1-\alpha_{2}}\log\left( \sum p_{j}^{\alpha_{2}}\right) \\ 
\vdots \\ 
\frac{1}{1-\alpha_{m}}\log\left( \sum p_{j}^{\alpha_{m}}\right) \\ 
\sum p_{j}%
\end{array}
\right) .  \label{open}
\end{equation}
The matrix of partial derivatives is%
\begin{equation*}
\left( 
\begin{array}{ccccc}
\frac{\alpha_{1}}{1-\alpha_{1}}\frac{p_{1}^{\alpha_{1}-1}}{\sum
p_{j}^{\alpha_{1}}} & \frac{\alpha_{1}}{1-\alpha_{1}}\frac{%
p_{2}^{\alpha_{1}-1}}{\sum p_{j}^{\alpha_{1}}} & \cdots & \frac{\alpha_{1}}{%
1-\alpha_{1}}\frac{p_{n-1}^{\alpha_{1}-1}}{\sum p_{j}^{\alpha_{1}}} & \frac{%
\alpha_{1}}{1-\alpha_{1}}\frac{p_{n}^{\alpha_{1}-1}}{\sum p_{j}^{\alpha_{1}}}
\\ 
\frac{\alpha_{2}}{1-\alpha_{2}}\frac{p_{1}^{\alpha_{2}-1}}{\sum
p_{j}^{\alpha_{2}}} & \frac{\alpha_{2}}{1-\alpha_{2}}\frac{%
p_{2}^{\alpha_{2}-1}}{\sum p_{j}^{\alpha_{2}}} & \cdots & \frac{\alpha_{2}}{%
1-\alpha_{2}}\frac{p_{n-1}^{\alpha_{2}-1}}{\sum p_{j}^{\alpha_{2}}} & \frac{%
\alpha_{2}}{1-\alpha_{2}}\frac{p_{n}^{\alpha_{2}-1}}{\sum p_{j}^{\alpha_{2}}}
\\ 
\vdots & \vdots & \ddots & \vdots & \vdots \\ 
\frac{\alpha_{m}}{1-\alpha_{m}}\frac{p_{1}^{\alpha_{m}-1}}{\sum
p_{j}^{\alpha_{m}}} & \frac{\alpha_{m}}{1-\alpha_{m}}\frac{%
p_{2}^{\alpha_{m}-1}}{\sum p_{j}^{\alpha_{m}}} & \cdots & \frac{\alpha_{m}}{%
1-\alpha_{m}}\frac{p_{n-1}^{\alpha_{m}-1}}{\sum p_{j}^{\alpha_{m}}} & \frac{%
\alpha_{m}}{1-\alpha_{m}}\frac{p_{n}^{\alpha_{m}-1}}{\sum p_{j}^{\alpha_{m}}}
\\ 
1 & 1 & \cdots & 1 & 1%
\end{array}
\right) .
\end{equation*}
If this matrix has rank }$m+1$ {\normalsize in a neighborhood of a point }$%
P= ${\normalsize $\left( p_{1},p_{2},...,p_{n}\right) $ then the map (\ref%
{open}) is open, i.e. it maps open sets into open sets and a neighborhood of 
}$P$ {\normalsize is mapped into a neighborhood of the image}.

{\normalsize Next we show that if $P$ has }$m+1$ {\normalsize different
point probabilities then $P$ is mapped into an interior point in the range.
Therefore, assume that $P$ has }$m+1$ {\normalsize different point
probabilities. For simplicity we may assume that these $m+1\,$different
point probabilities are the first ones and that }$0<p_{1}<p_{2}<%
\cdots<p_{m+1}.$ {\normalsize Then}%
%\begin{multline}
\begin{align}
&
 \left\vert 
\begin{array}{ccccc}
\frac{\alpha_{1}}{1-\alpha_{1}}\frac{p_{1}^{\alpha_{1}-1}}{\sum
p_{j}^{\alpha_{1}}} & \frac{\alpha_{1}}{1-\alpha_{1}}\frac{%
p_{2}^{\alpha_{1}-1}}{\sum p_{j}^{\alpha_{1}}} & \cdots & \frac{\alpha_{1}}{%
1-\alpha_{1}}\frac{p_{m}^{\alpha_{1}-1}}{\sum p_{j}^{\alpha_{1}}} & \frac{%
\alpha_{1}}{1-\alpha_{1}}\frac{p_{m+1}^{\alpha_{1}-1}}{\sum
p_{j}^{\alpha_{1}}} \\ 
\frac{\alpha_{2}}{1-\alpha_{2}}\frac{p_{1}^{\alpha_{2}-1}}{\sum
p_{j}^{\alpha_{2}}} & \frac{\alpha_{2}}{1-\alpha_{2}}\frac{%
p_{2}^{\alpha_{2}-1}}{\sum p_{j}^{\alpha_{2}}} & \cdots & \frac{\alpha_{2}}{%
1-\alpha_{2}}\frac{p_{m}^{\alpha_{2}-1}}{\sum p_{j}^{\alpha_{2}}} & \frac{%
\alpha_{2}}{1-\alpha_{2}}\frac{p_{m+1}^{\alpha_{2}-1}}{\sum
p_{j}^{\alpha_{2}}} \\ 
\vdots & \vdots & \ddots & \vdots & \vdots \\ 
\frac{\alpha_{m}}{1-\alpha_{m}}\frac{p_{1}^{\alpha_{m}-1}}{\sum
p_{j}^{\alpha_{m}}} & \frac{\alpha_{m}}{1-\alpha_{m}}\frac{%
p_{2}^{\alpha_{m}-1}}{\sum p_{j}^{\alpha_{m}}} & \cdots & \frac{\alpha_{m}}{%
1-\alpha_{m}}\frac{p_{m}^{\alpha_{m}-1}}{\sum p_{j}^{\alpha_{m}}} & \frac{%
\alpha_{m}}{1-\alpha_{m}}\frac{p_{m+1}^{\alpha_{m}-1}}{\sum
p_{j}^{\alpha_{m}}} \\ 
1 & 1 & \cdots & 1 & 1%
\end{array}
\right\vert  \label{determinant} \\
&
 =\left( 
\prod \limits_{i=1}^{m}%
\frac{\alpha_{i}}{1-\alpha_{i}}\cdot\prod
\limits_{i=1}^{m}\frac{1}{\sum_{j}p_{j}^{\alpha_{1}}}
\right) 
\left\vert 
\begin{array}{ccccc}
p_{1}^{\alpha_{1}-1} & p_{2}^{\alpha_{1}-1} & \cdots & p_{m}^{\alpha_{1}-1}
& p_{m+1}^{\alpha_{1}-1} \\ 
p_{1}^{\alpha_{2}-1} & p_{2}^{\alpha_{2}-1} & \cdots & p_{m}^{\alpha_{2}-1}
& p_{m+1}^{\alpha_{2}-1} \\ 
\vdots & \vdots & \ddots & \vdots & \vdots \\ 
p_{1}^{\alpha_{m}-1} & p_{2}^{\alpha_{m}-1} & \cdots & p_{m}^{\alpha_{m}-1}
& p_{m+1}^{\alpha_{m}-1} \\ 
1 & 1 & \cdots & 1 & 1%
\end{array}
\right\vert \notag
\end{align}
%\end{multline}
{\normalsize Note that the last row can be written as }$\left( 
\begin{array}{ccccc}
p_{1}^{\alpha-1} & p_{2}^{\alpha-1} & \cdots & p_{m}^{\alpha-1} & 
p_{m+1}^{\alpha-1}%
\end{array}
\right) ${\normalsize with }$\alpha=1.$ {\normalsize The last determinant is
a generalization of the Vandermonde determinant. Like a Vandermonde
determinant, it is non-zero if and only if the entries are different, which
is the next we have to prove.}

\begin{lemma}
Assume that $0<x_{1}\leq x_{2}\leq\cdots\leq x_{\ell}$ and $%
\beta_{1}<\beta_{2}<\cdots<\beta_{\ell}.$ Then the generalized Vandermonde
determinant%
\begin{equation*}
\det\left( \left( x_{i}^{\beta_{j}}\right) _{i,j=1,2,\cdots\ell}\right)
=\left\vert 
\begin{array}{cccc}
x_{1}^{\beta_{1}} & x_{1}^{\beta_{1}} & \cdots & x_{\ell}^{\beta_{1}} \\ 
x_{1}^{\beta_{2}} & x_{2}^{\beta_{2}} & \cdots & x_{\ell}^{\beta_{2}} \\ 
\vdots & \vdots & \ddots & \vdots \\ 
x_{1}^{\beta_{\ell}} & x_{2}^{\beta_{\ell}} & \cdots & x_{\ell}^{\beta_{%
\ell}}%
\end{array}
\right\vert
\end{equation*}
is non-negative. It is zero if and only if there exists $j\in \left\lbrace
1,2,\cdots\ell-1 \right\rbrace$
such that $x_{j}=x_{j+1}.$
\end{lemma}

\bigskip\textbf{Proof} {\normalsize The proof is by induction in} $\ell.$ 
{\normalsize For} $\ell=1$ {\normalsize the generalized Vandermonde
determinant is obviously positive. Assume that the result holds for} $%
\ell=k-1.$ {\normalsize We have to prove it for} $\ell=k.$ {\normalsize %
First we have} 
\begin{equation*}
\left\vert 
\begin{array}{cccc}
x_{1}^{\beta_{1}} & x_{2}^{\beta_{1}} & \cdots & x_{k}^{\beta_{1}} \\ 
x_{1}^{\beta_{2}} & x_{2}^{\beta_{2}} & \cdots & x_{k}^{\beta_{2}} \\ 
\vdots & \vdots & \ddots & \vdots \\ 
x_{1}^{\beta_{k}} & x_{2}^{\beta_{k}} & \cdots & x_{k}^{\beta_{k}}%
\end{array}
\right\vert =\prod \limits_{j=1}^{k}x_{j}^{\beta_{1}}\left\vert 
\begin{array}{cccc}
1 & 1 & \cdots & 1 \\ 
x_{1}^{\beta_{2}-\beta_{1}} & x_{2}^{\beta_{2}-\beta_{1}} & \cdots & 
x_{k}^{\beta_{2}-\beta_{1}} \\ 
\vdots & \vdots & \ddots & \vdots \\ 
x_{1}^{\beta_{k}-\beta_{1}} & x_{2}^{\beta_{k}-\beta_{1}} & \cdots & 
x_{k}^{\beta_{k}}%
\end{array}
\right\vert .
\end{equation*}
{\normalsize Therefore without loss of generality we may assume that} $%
\beta_{1}=0.$ {\normalsize Therefore we have to prove that} 
\begin{equation*}
\left\vert 
\begin{array}{ccccc}
1 & 1 & \cdots & 1 & 1 \\ 
x_{1}^{\beta_{2}} & x_{2}^{\beta_{2}} & \cdots & x_{k-1}^{\beta_{2}} & 
x_{k}^{\beta_{2}} \\ 
x_{1}^{\beta_{3}} & x_{2}^{\beta_{3}} & \cdots & x_{k-1}^{\beta_{3}} & 
x_{k}^{\beta_{3}} \\ 
\vdots & \vdots & \ddots & \vdots & \vdots \\ 
x_{1}^{\beta_{k}} & x_{2}^{\beta_{k}} & \cdots & x_{k-1}^{\beta_{k}} & 
x_{k}^{\beta_{k}}%
\end{array}
\right\vert
\end{equation*}
{\normalsize is non-negative. If} $x_{k}=x_{k-1}$ {\normalsize the last two
columns are identical and determinant is zero so it is sufficient to prove
that the partial derivative with respect to} $x_{k}$ {\normalsize is
non-negative. The partial derivative is}%
\begin{equation*}
\left\vert 
\begin{array}{ccccc}
1 & 1 & \cdots & 1 & 0 \\ 
x_{1}^{\beta_{2}} & x_{2}^{\beta_{2}} & \cdots & x_{3}^{\beta_{2}} & \beta
_{2}x_{k}^{\beta_{2}-1} \\ 
x_{1}^{\beta_{3}} & x_{2}^{\beta_{3}} & \cdots & x_{3}^{\beta_{3}} & \beta
_{3}x_{k}^{\beta_{3}-1} \\ 
\vdots & \vdots & \ddots & \vdots & \vdots \\ 
x_{1}^{\beta_{k}} & x_{2}^{\beta_{k}} & \cdots & x_{3}^{\beta_{k}} & \beta
_{k}x_{k}^{\beta_{2}-1}%
\end{array}
\right\vert .
\end{equation*}
{\normalsize Similarly we may take partial derivatives with respect to} $%
x_{k-1},x_{k-2},\cdots,x_{3}$ {\normalsize and} $x_{2}$ {\normalsize and get}%
\begin{multline*}
\left\vert 
\begin{array}{ccccc}
1 & 0 & 0 & \cdots & 0 \\ 
x_{1}^{\beta_{2}} & \beta_{2}x_{2}^{\beta_{2}-1} & \beta_{2}x_{3}^{\beta
_{2}-1} & \cdots & \beta_{2}x_{k}^{\beta_{2}-1} \\ 
x_{1}^{\beta_{3}} & \beta_{3}x_{2}^{\beta_{3}-1} & \beta_{3}x_{3}^{\beta
_{3}-1} & \cdots & \beta_{3}x_{k}^{\beta_{3}-1} \\ 
\vdots & \vdots & \vdots & \ddots & \vdots \\ 
x_{1}^{\beta_{k}} & \beta_{k}x_{2}^{\beta_{2}-1} & \beta_{k}x_{3}^{\beta
_{2}-1} & \cdots & \beta_{k}x_{k}^{\beta_{2}-1}%
\end{array}
\right\vert \\
=\left\vert 
\begin{array}{cccc}
\beta_{2}x_{2}^{\beta_{2}-1} & \beta_{2}x_{3}^{\beta_{2}-1} & \cdots & 
\beta_{2}x_{k}^{\beta_{2}-1} \\ 
\beta_{3}x_{2}^{\beta_{3}-1} & \beta_{3}x_{3}^{\beta_{3}-1} & \cdots & 
\beta_{3}x_{m}^{\beta_{3}-1} \\ 
\vdots & \vdots & \ddots & \vdots \\ 
\beta_{k}x_{2}^{\beta_{2}-1} & \beta_{k}x_{3}^{\beta_{2}-1} & \cdots & 
\beta_{k}x_{k}^{\beta_{2}-1}%
\end{array}
\right\vert \\
=\prod \limits_{j=1}^{k}\beta_{j}\left\vert 
\begin{array}{cccc}
x_{2}^{\beta_{2}-1} & x_{3}^{\beta_{2}-1} & \cdots & x_{k}^{\beta_{2}-1} \\ 
x_{2}^{\beta_{3}-1} & x_{3}^{\beta_{3}-1} & \cdots & x_{k}^{\beta_{3}-1} \\ 
\vdots & \vdots & \ddots & \vdots \\ 
x_{2}^{\beta_{2}-1} & x_{3}^{\beta_{2}-1} & \cdots & x_{k}^{\beta_{2}-1}%
\end{array}
\right\vert .
\end{multline*}
{\normalsize This is non-negative according to the induction hypothesis.}$\
\ \ \blacksquare$

{\normalsize We see that if }$0<\alpha_{1}<\cdots<\alpha_{m}<1$ {\normalsize %
then the determinant (\ref{determinant}) is positive. It is easy to check
that this is also the case with the relaxed condition }$0<\alpha
_{1}<\cdots<\alpha_{m}.$

{\normalsize The R\'{e}nyi entropies are symmetric in their entries.
Therefore we may restrict our attention to probability vectors with
increasing entries, i.e. }$0\leq p_{1}\leq p_{2}\leq\cdots\leq p_{m+1}.$ 
{\normalsize The extreme points in the set of ordered probability vectors
are the uniform distributions. Let }$U_{k}$ {\normalsize denote the uniform
distribution }$\left( 0,0,\cdots,0,\frac{1}{k},\frac{1}{k},\cdots,\frac{1}{k}%
\right) .$ {\normalsize Let }$k_{1},k_{2},\ldots,k_{\ell}$ {\normalsize be a
sequence of different numbers in }$\left\{ 1,2,\cdots,n\right\} .$ 
{\normalsize Then the simplex formed by convex combinations of }$%
U_{k_{1}},U_{k_{2}},\ldots,U_{k_{\ell}}$ {\normalsize will shall be denoted }%
$\Delta_{k_{1},k_{2},\cdots,k_{\ell}}$ {\normalsize and be given an
orientation according to the sequence }$U_{k_{1}},U_{k_{2}},\ldots,U_{k_{%
\ell }}.$ {\normalsize Observe that if }$k_{1}>k_{2}>\ldots>k_{m+1}$ 
{\normalsize then the mapping }$\Delta_{k_{1},k_{2},\cdots,k_{m}}\rightarrow%
\mathbb{R}^{m}$ {\normalsize defined by}%
\begin{equation*}
P\rightarrow\left( 
\begin{array}{c}
\frac{1}{1-\alpha_{1}}\log\left( \sum p_{j}^{\alpha_{1}}\right) \\ 
\frac{1}{1-\alpha_{2}}\log\left( \sum p_{j}^{\alpha_{2}}\right) \\ 
\vdots \\ 
\frac{1}{1-\alpha_{m}}\log\left( \sum p_{j}^{\alpha_{m}}\right)%
\end{array}
\right)
\end{equation*}
{\normalsize has positive orientation if }$0<\alpha_{1}<\alpha_{2}<\cdots<%
\alpha_{m}.$

\section{{\protect\normalsize \hs JOINT RANGE OF TWO R\'{E}NYI ENTROPIES}}

{\normalsize First we consider distributions on a set with }$n$ {\normalsize %
elements. We determine the joint range of }$H_{\alpha_{1}}$ {\normalsize and 
}$H_{\alpha_{2}}$ {\normalsize where we assume that }$0<\alpha_{1}<%
\alpha_{2}.$ {\normalsize First we shall also assume that }$%
\alpha_{1},\alpha_{2}\in\left] 0;\infty\right[ \backslash\left\{ 1\right\} .$
{\normalsize Let }$\Phi$ {\normalsize denote the map }%
\begin{equation*}
P\rightarrow\left( 
\begin{array}{c}
H_{\alpha_{1}}\left( P\right) \\ 
H_{\alpha_{2}}\left( P\right)%
\end{array}
\right) .
\end{equation*}

{\normalsize Assume that }$k_{1}>k_{2}>k_{3}.$ {\normalsize Then }$%
\Phi\left( U_{k_{j}}\right) $ {\normalsize lies on the diagonal }$\left\{
\left( x,x\right) :x\geq0\right\} ,$ {\normalsize and these points are
ordered, }%
\begin{equation*}
H_{\alpha}\left( U_{k_{1}}\right) >H_{\alpha}\left( U_{k_{2}}\right)
>H_{\alpha}\left( U_{k_{3}}\right)
\end{equation*}
{\normalsize where }$\alpha=\alpha_{1}$ {\normalsize or }$\alpha=\alpha_{2}.$
{\normalsize We know that }$H_{a_{1}}\left( P\right) \geq
H_{\alpha_{2}}\left( P\right) $ {\normalsize with equality if and only if }$%
P $ {\normalsize is a uniform distribution. Therefore }$\Phi$ {\normalsize %
restricted to }$\Delta_{k_{1},k_{2},k_{3}}$ {\normalsize must preserve
orientation. We know that }$\Phi$ {\normalsize maps inner points of }$%
\Delta_{k_{1},k_{2},k_{3}}$ {\normalsize into inner points of the range of }$%
\Phi$ {\normalsize so boundary points of the range of }$\Phi$ {\normalsize %
must have preimages that are boundary points of }$\Delta
_{k_{1},k_{2},k_{3}}.$ {\normalsize We follow the conventions from homology
theory and calculate the boundary with orientation. The boundary of }$%
\Phi\left( \Delta_{k_{1},k_{2},k_{3}}\right) $ {\normalsize is }%
\begin{align*}
\partial\Phi\left( \Delta_{k_{1},k_{2},k_{3}}\right) & =\Phi \partial\left(
\Delta_{k_{1},k_{2},k_{3}}\right) \\
& =\Phi\left(
\Delta_{k_{2},k_{3}}-\Delta_{k_{1},k_{3}}+\Delta_{k_{1},k_{2}}\right) \\
& =\Phi\left(
\Delta_{k_{1},k_{2}}+\Delta_{k_{2},k_{3}}+\Delta_{k_{3},k_{1}}\right) ,
\end{align*}
{\normalsize which is just another way of writing the closed curve from }$%
U_{k_{1}}$ {\normalsize to }$U_{k_{2}}$ {\normalsize to }$U_{k_{3}}$ 
{\normalsize and back to }$U_{k_{1}}.$ {\normalsize Therefore any point on
the boundary of the range of }$\Phi$ {\normalsize must be the image of a
mixture of two uniform distributions.}

{\normalsize Assume that }$k_{1}>k_{2}>k_{3}>k_{4}.$ {\normalsize Then the
simplices }$\Delta_{k_{1},k_{2},k_{3}}$ {\normalsize and }$%
\Delta_{k_{1},k_{3},k_{4}}$ {\normalsize are both positively oriented and}%
\begin{align*}
\partial\Phi\left(
\Delta_{k_{1},k_{2},k_{3}}+\Delta_{k_{1},k_{3},k_{4}}\right) &
=\Phi\partial\left( \Delta_{k_{1},k_{2},k_{3}}+\Delta
_{k_{1},k_{3},k_{4}}\right) \\
& =\Phi\left( 
\begin{array}{c}
\partial\Delta_{k_{1},k_{2},k_{3}} \\ 
+\partial\Delta_{k_{1},k_{3},k_{4}}%
\end{array}
\right) \\
& =\Phi\left( 
\begin{array}{c}
\Delta_{k_{2},k_{3}}-\Delta_{k_{1},k_{3}}+\Delta_{k_{1},k_{2}} \\ 
+\Delta_{k_{3},k_{4}}-\Delta_{k_{1},k_{4}}+\Delta_{k_{3},k_{4}}%
\end{array}
\right) \\
& =\Phi\left( 
\begin{array}{c}
\Delta_{k_{2},k_{3}}-\Delta_{k_{1},k_{3}}+\Delta_{k_{1},k_{2}} \\ 
+\Delta_{k_{3},k_{4}}-\Delta_{k_{1},k_{4}}+\Delta_{k_{1},k_{3}}%
\end{array}
\right) \\
& =\Phi\left(
\Delta_{k_{1},k_{2}}+\Delta_{k_{2},k_{3}}+\Delta_{k_{3},k_{4}}+%
\Delta_{k_{4},k_{1}}\right) .
\end{align*}
{\normalsize We see that }$\Phi\left( \Delta_{k_{1},k_{3}}\right) $ 
{\normalsize does not contribute to the boundary of }%
\begin{equation*}
\partial\Phi\left(
\Delta_{k_{1},k_{2},k_{3}}+\Delta_{k_{1},k_{3},k_{4}}\right) .
\end{equation*}
{\normalsize Similarly }$\Phi\left( \Delta_{k_{2},k_{4}}\right) $ 
{\normalsize does not contribute to the boundary. We may formulate this
result as }$\Delta_{a,b}$ {\normalsize does not contribute to the range if
it is a diagonal in a quadruple. The non-diagonal simplices are }$\Delta
_{n,n-1},\Delta_{n-1,n-2},\cdots,\Delta_{2,1}${\normalsize and }$%
\Delta_{1,n} ${\normalsize . These form a closed curve }%
\begin{equation*}
\Delta_{n,n-1}+\Delta_{n-1,n-2}+\cdots+\Delta_{2,1}+\Delta_{1,n}
\end{equation*}
{\normalsize and the boundary is the image of this curve, i.e.}%
\begin{equation*}
\Phi\left( \Delta_{n,n-1}+\Delta_{n-1,n-2}+\cdots+\Delta_{2,1}+\Delta
_{1,n}\right) ,
\end{equation*}
{\normalsize This result easily extends to the cases where one or more of
the orders equal }${1}$ {\normalsize or }$\infty.$ {\normalsize The upper
bound does not depend on} $n$ {\normalsize so we get the following theorem.}

\begin{theorem}
{\normalsize Assume }$0<\alpha_{1}<\alpha_{2}.$ {\normalsize Then the upper
bound on $H_{\alpha_{2}}\left( P\right) $ given $H_{\alpha_{1}}\left(
P\right) $ is attained by a mixture of uniform distributions on $k$ and $k+1$
points where $k$ is determined by the condition $\log k\leq
H_{\alpha_{1}}\left( P\right) <\log\left( k+1\right) .$ }
\end{theorem}

{\normalsize For distributions on set with }$n$ {\normalsize elements we
also get a tight lower bound, but if we have no restriction on }$n$ 
{\normalsize the situation is a little more complicated.}

\begin{theorem}
{\normalsize Assume }$0<\alpha_{1}<\alpha_{2}.$ {\normalsize If }$P$ 
{\normalsize is a distribution on a set with }$n$ {\normalsize elements and }%
$H_{\alpha_{1}}\left( P\right) $ {\normalsize is fixed then a lower bound on 
}$H_{\alpha_{2}}$ {\normalsize is attained for a mixture of the uniform
distributions }$U_{1}$ {\normalsize and }$U_{n}.$ {\normalsize If no
restriction on }$n$ {\normalsize is given and if }$H_{\alpha_{1}}\left(
P\right) >0$ {\normalsize is fixed then a tight lower bound on }$%
H_{\alpha_{2}}\left( P\right) $ {\normalsize is given by}%
\begin{equation*}
H_{a_{2}}\left( P\right) >\left\{ 
\begin{array}{cc}
0, & \text{if }\alpha_{1}\leq1; \\ 
\frac{\alpha_{2}}{\alpha_{2}-1}\frac{\alpha_{1}-1}{\alpha_{1}}%
H_{\alpha_{1}}\left( P\right) , & \text{if }\alpha_{1}>1.%
\end{array}
\right.
\end{equation*}
\end{theorem}

\textbf{Proof} {\normalsize If we have no restriction on }$n$ {\normalsize %
then the range is }%
\begin{equation*}
\bigoplus \limits_{n=2}^{\infty}\Phi\left( \Delta_{n+1,n,1}\right) .
\end{equation*}
{\normalsize So we just have to determine the asymptotics of }$\Phi\left(
\Delta_{n,1}\right) .$ {\normalsize The curve }$\Delta_{1,n}$ {\normalsize %
has the parametrization }$P_{t}=\left( \frac{t}{n},\frac{t}{n},\cdots,\frac{t%
}{n},\frac{t}{n}+1-t\right) ,~$$t\in\left[ 0;1\right] .$ {\normalsize %
Therefore the curve }$\Phi\left( \Delta_{n,1}\right) $ {\normalsize has the
parametrization}%
\begin{equation*}
\left( 
\begin{array}{c}
\frac{1}{1-\alpha_{1}}\log\left( \left( n-1\right) \left( \frac{t}{n}\right)
^{\alpha_{1}}+\left( \frac{t}{n}+1-t\right) ^{\alpha_{1}}\right) \\ 
\frac{1}{1-\alpha_{2}}\log\left( \left( n-1\right) \left( \frac{t}{n}\right)
^{\alpha_{2}}+\left( \frac{t}{n}+1-t\right) ^{\alpha_{2}}\right)%
\end{array}
\right) .
\end{equation*}
{\normalsize We have to study the asymptotics of this curve for }$n$ 
{\normalsize tending to infinity. There are several cases and they need
separate analysis.}

\begin{description}
\item[{\protect\normalsize Case }$\protect\alpha_{1}>1.$ ] {\normalsize We
also have }$\alpha_{2}>1$ {\normalsize so for a fixed value of }$t$ 
{\normalsize we get}%
\begin{equation*}
\left( 
\begin{array}{c}
\frac{1}{1-\alpha_{1}}\log\left( \left( n-1\right) \left( \frac{t}{n}\right)
^{\alpha_{1}}+\left( \frac{t}{n}+1-t\right) ^{\alpha_{1}}\right) \\ 
\frac{1}{1-\alpha_{2}}\log\left( \left( n-1\right) \left( \frac{t}{n}\right)
^{\alpha_{2}}+\left( \frac{t}{n}+1-t\right) ^{\alpha_{2}}\right)%
\end{array}
\right) \rightarrow\left( 
\begin{array}{c}
\frac{\alpha_{1}}{1-\alpha_{1}}\log\left( 1-t\right) \\ 
\frac{\alpha_{2}}{1-\alpha_{2}}\log\left( 1-t\right)%
\end{array}
\right)
\end{equation*}
{\normalsize for $n$ tending to infinity. Hence the straight line with slope 
}$\frac{\alpha_{2}}{\alpha_{2}-1}\frac{\alpha_{1}-1}{\alpha_{1}}$ 
{\normalsize is the boundary of the range. }

\item[Case $\protect\alpha_{2}\geq1$ and $\protect\alpha_{1}\leq1.$] 
{\normalsize First we assume that }$\alpha_{1}<1.$ {\normalsize For a fixed
value of the parameter }$t$ {\normalsize the R\'{e}nyi entropy }$H_{a_{2}}$ 
{\normalsize tends to a constant as above but }$H_{\alpha_{1}}$ {\normalsize %
tends to infinity. For a fixed value of }$H_{\alpha_{1}}\left( P\right) >0$ 
{\normalsize the lower bound }$H_{\alpha_{2}}\left( P\right) >0$ 
{\normalsize is tight. This bound is also tight for }$\alpha_{1}=1$ 
{\normalsize and can be obtained by letting }$\alpha_{1}$ {\normalsize tend
to 1 from above or below. }

\item[{\protect\normalsize Case }$0<\protect\alpha_{2}\leq1.$ ] {\normalsize %
First assume that }$\alpha_{2}<1.$ {\normalsize If }$t=n^{1-1/\alpha_{2}}$ 
{\normalsize then}%
\begin{align*}
& \left( 
\begin{array}{c}
\frac{1}{1-\alpha_{1}}\log\left( \left( n-1\right) \left( \frac{t}{n}\right)
^{\alpha_{1}}+\left( \frac{t}{n}+1-t\right) ^{\alpha_{1}}\right) \\ 
\frac{1}{1-\alpha_{2}}\log\left( \left( n-1\right) \left( \frac{t}{n}\right)
^{\alpha_{2}}+\left( \frac{t}{n}+1-t\right) ^{\alpha_{2}}\right)%
\end{array}
\right) \\
& =\left( 
\begin{array}{c}
\frac{1}{1-\alpha_{1}}\log\left( n^{-\frac{\alpha_{1}}{\alpha_{2}}}\cdot 
\frac{n-1}{n}+\left( n^{-1/\alpha_{2}}+1-n^{1-1/\alpha_{2}}\right)
^{\alpha_{1}}\right) \\ 
\frac{1}{1-\alpha_{2}}\log\left( \frac{n-1}{n}+\left(
n^{-1/\alpha_{2}}+1-n^{1-1/\alpha_{2}}\right) ^{\alpha_{2}}\right)%
\end{array}
\right) .
\end{align*}
{\normalsize We see that the second coordinate tends to }$\frac{1}{%
1-\alpha_{2}}\log2,$ {\normalsize while the first coordinate tends to }$%
\infty.$ {\normalsize Therefore for a fixed value of }$H_{\alpha_{1}}\left(
P\right) >0$ {\normalsize the lower bound }$H_{\alpha_{2}}\left( P\right) >0$
{\normalsize is tight. Tightness of this bound also holds for }$\alpha
_{2}=1 ${\normalsize , which can be seen by letting }$\alpha_{2}$ 
{\normalsize tend }$1$ {\normalsize from above or from below.}$\ \ \
\blacksquare$
\end{description}

\section{{\protect\normalsize \hs JOINT RANGE OF\ THREE R\'{E}NYI ENTROPIES}}

{\normalsize Determining the range of three R\'{e}nyi entropies is done in
the same way as in the previous section. We consider the map }$\Psi$ 
{\normalsize given by}%
\begin{equation*}
P\rightarrow\left( 
\begin{array}{c}
H_{\alpha_{1}}\left( P\right) \\ 
H_{\alpha_{2}}\left( P\right) \\ 
H_{\alpha_{3}}\left( P\right)%
\end{array}
\right) .
\end{equation*}
{\normalsize First we consider the situation where the domain consist of
distributions on }$n$ {\normalsize points. Boundary points of }$\Psi$ 
{\normalsize must be images of mixtures of three uniform distributions. If }$%
n>m>\ell>k>1$ {\normalsize then the restriction of }$\Psi$ {\normalsize to
the simplices }$\Delta_{n,m,\ell,k}$ {\normalsize or to }$%
\Delta_{m,\ell,k,1} $ {\normalsize conserves orientation. Therefore }%
\begin{align*}
\partial\Psi\left( \Delta_{n,m,\ell,k}+\Delta_{m,\ell,k,1}\right) &
=\partial\Psi\left( \partial\Delta_{n,m,\ell,k}+\partial\Delta_{m,\ell
,k,1}\right) \\
& =\partial\Psi\left( 
\begin{array}{c}
\Delta_{m,\ell,k}-\Delta_{n,\ell,k}+\Delta_{n,m,k}-\Delta_{n,m,\ell} \\ 
+\Delta_{\ell,k,1}-\Delta_{m,k,1}+\Delta_{m,\ell,1}-\Delta_{m,\ell,k}%
\end{array}
\right) \\
& =\partial\Psi\left( 
\begin{array}{c}
-\Delta_{n,\ell,k}+\Delta_{n,m,k}-\Delta_{n,m,\ell} \\ 
+\Delta_{\ell,k,1}-\Delta_{m,k,1}+\Delta_{m,\ell,1}%
\end{array}
\right) .
\end{align*}
{\normalsize We see that }$\Delta_{m,\ell,k}$ {\normalsize gives no
contribution to the boundary and therefore only simplices }$%
\Delta_{m,\ell,k} $ {\normalsize with either }$m=n$ {\normalsize or }$k=1$ 
{\normalsize give a contributions to the boundary.}

{\normalsize If }$n>m>\ell >k>1$ {\normalsize then the restriction of }$\Psi 
$ {\normalsize to the simplices }$\Delta _{n,m,k,1}$ {\normalsize or to }$%
\Delta _{m,\ell ,k,1}$ {\normalsize conserves orientation. Therefore}%
\begin{align*}
\partial \Psi \left( \Delta _{n,m,k,1}+\Delta _{m,\ell ,k,1}\right) &
=\partial \Psi \left( \partial \Delta _{n,m,k,1}+\partial \Delta _{m,\ell
,k,1}\right) \\
& =\partial \Psi \left( 
\begin{array}{c}
\Delta _{m,k,1}-\Delta _{n,k,1}+\Delta _{n,m,1}-\Delta _{n,m,k} \\ 
+\Delta _{\ell ,k,1}-\Delta _{m,k,1}+\Delta _{m,\ell ,1}-\Delta _{m,\ell ,k}%
\end{array}%
\right) \\
& =\partial \Psi \left( 
\begin{array}{c}
-\Delta _{n,k,1}+\Delta _{n,m,1}-\Delta _{n,m,k} \\ 
+\Delta _{\ell ,k,1}+\Delta _{m,\ell ,1}-\Delta _{m,\ell ,k}%
\end{array}%
\right) .
\end{align*}%
{\normalsize We see that the simplex }$\Delta _{m,k,1}$ {\normalsize gives
no contribution to the boundary of the range of }$\Psi .$ {\normalsize So if 
}$m<n$ {\normalsize the simplex }$\Delta _{m,k,1}$ {\normalsize can only
give a contribution to the boundary if there exist no natural number }$\ell $
{\normalsize such that }$m>\ell >k,$ {\normalsize i.e. }$k=m-1.$ 
{\normalsize In the same way we can show that a simplex of the form }$\Delta
_{n,m,\ell }$ {\normalsize will only contribute to the boundary if }$\ell
=m-1$ {\normalsize and that a simplex }$\Delta _{n,m,1}$ {\normalsize only
contributes if }$m=n-1$ {\normalsize or if }$m=2.$ {\normalsize Thus the
boundary of the range consist of images of the simplices }$\Delta _{m,m-1,1}$
{\normalsize and of the form }$\Delta _{n,m,m-1}.$ {\normalsize Here we
notice that }%
\begin{multline*}
%\begin{align*}
\partial \left( \bigoplus\limits_{m=3}^{n}\Delta
_{m,m-1,1}-\bigoplus\limits_{m=2}^{n-1}\Delta _{n,m,m-1}\right) 
=\bigoplus\limits_{m=3}^{n}\partial \Delta
_{m,m-1,1}-\bigoplus\limits_{m=2}^{n-1}\partial \Delta _{n,m,m-1} \\
 =\bigoplus\limits_{m=3}^{n}\left( \Delta _{m-1,1}-\Delta _{m,1}+\Delta
_{m,m-1}\right) \\
 -\bigoplus\limits_{m=2}^{n-1}\left( \Delta _{m,m-1}-\Delta
_{n,m-1}+\Delta _{n,m}\right) %\\
 =0,
%\end{align*}%
\end{multline*}

{\normalsize so that }%
\begin{equation*}
\bigoplus\limits_{m=3}^{n}\Delta
_{m,m-1,1}-\bigoplus\limits_{m=2}^{n-1}\Delta _{n,m,m-1}
\end{equation*}%
{\normalsize is a closed surface and that the range of }$\Psi $ {\normalsize %
has the image of this surface as boundary.

\begin{figure}[htp]
\centering
\includegraphics[width=2in]{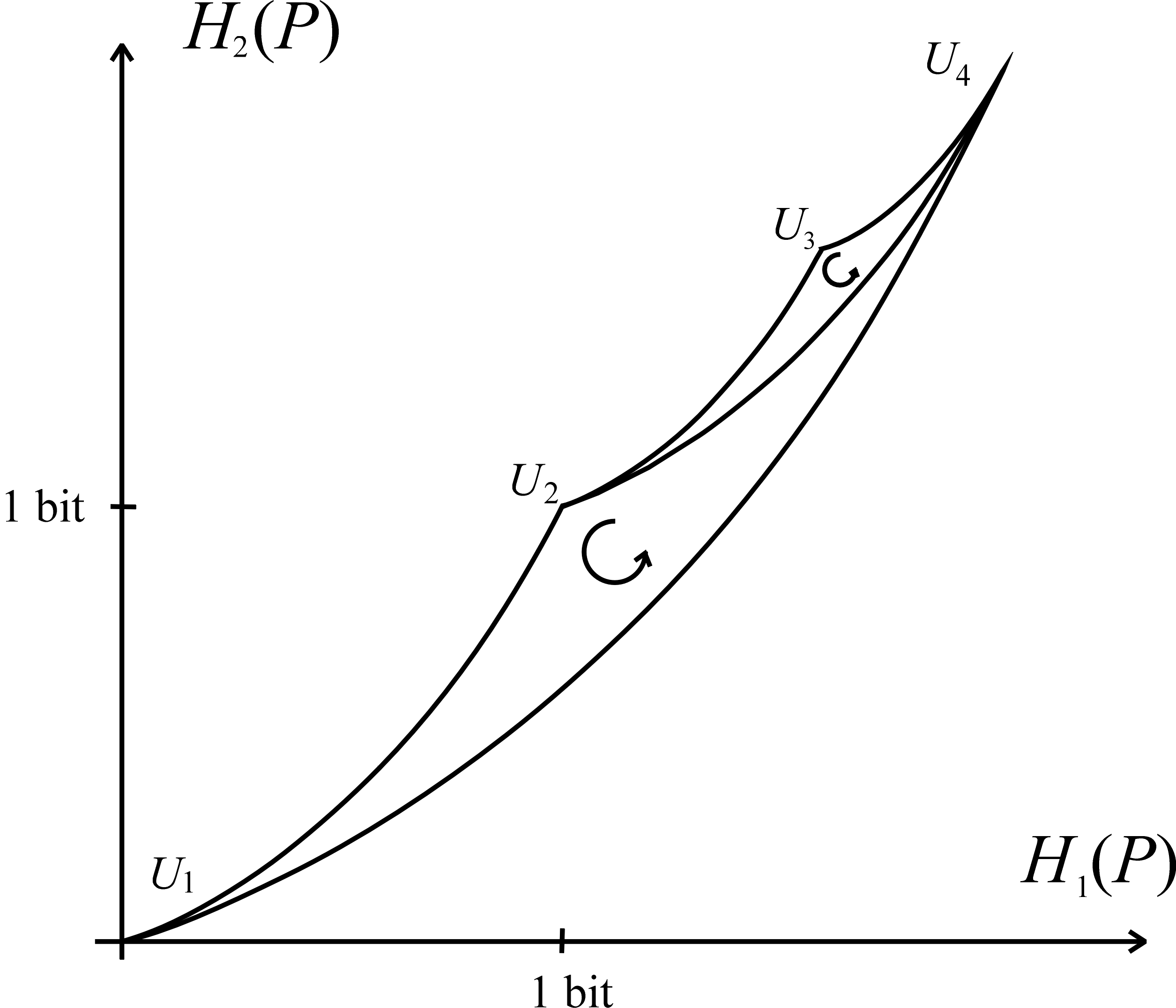}\includegraphics[width=2in]{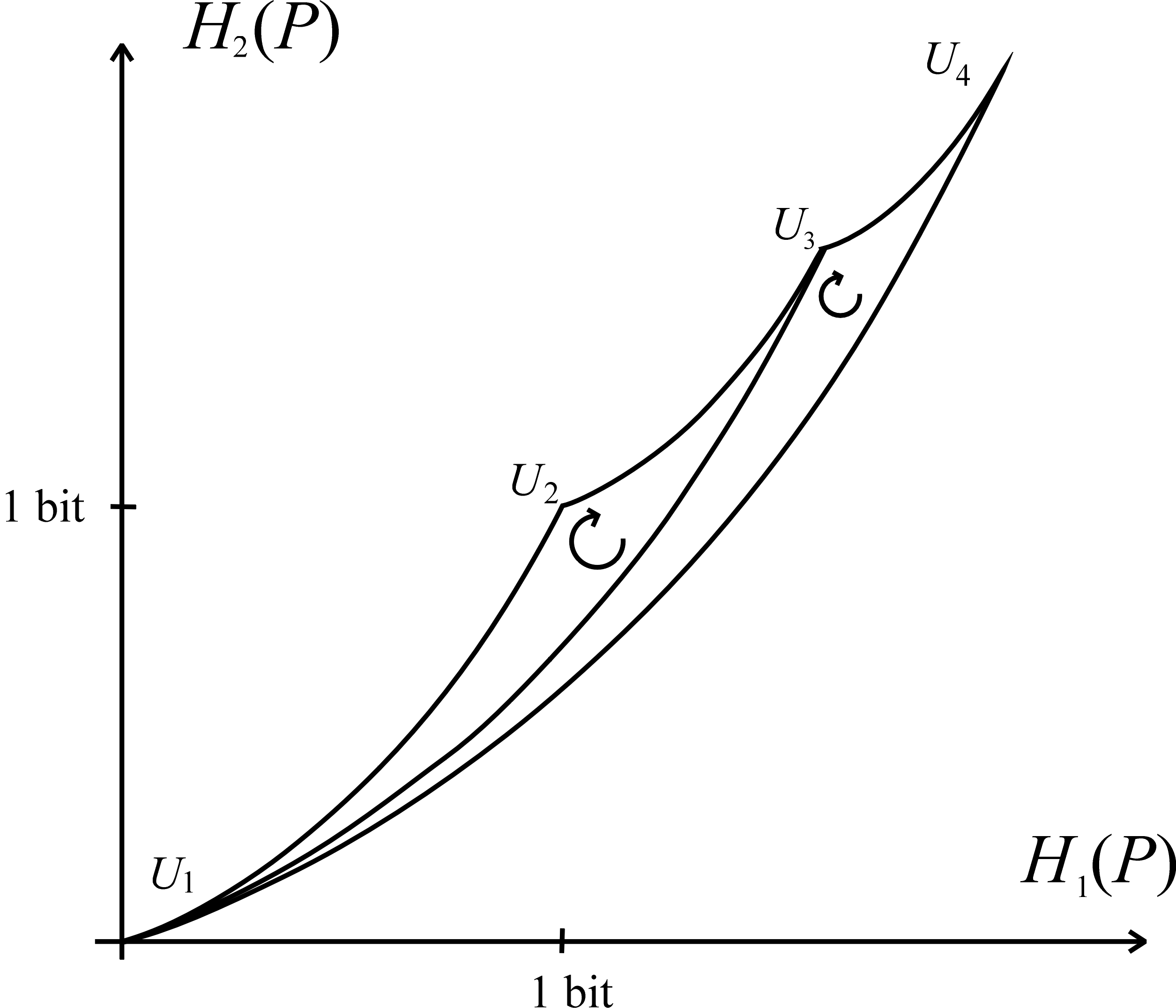}
\caption{The left diagram illustrates the range of $\Delta _{4,2,1}$ and $\Delta _{4,3,2}$ with orientation ($%
n=4$). The range $\Psi $ applied to these simplices give lower bounds on $%
H_{3}.$ The right diagram illustrates the range of $\Delta _{3,1,2}$ and $\Delta _{4,1,3}$ with orientation
($n=4$). The range $\Psi $ applied to these simplices give upper bounds on $%
H_{3}.$}\label{Fig1}
\end{figure}

%\FRAME{ftbpFU}{3.0234in}{2.5936in}{%
%0pt}{\Qcb{Range of $\Delta _{3,1,2}$ and $\Delta _{4,1,3}$ with orientation
%( $n=4$). The range $\Psi $ applied to these simplices give upper bounds on $%
%H_{3}.$}}{\Qlb{Fig1}}{lower.jpg}{\special{language "Scientific Word";type
%"GRAPHIC";maintain-aspect-ratio TRUE;display "USEDEF";valid_file "F";width
%3.0234in;height 2.5936in;depth 0pt;original-width 8.3662in;original-height
%7.1736in;cropleft "0";croptop "1";cropright "1";cropbottom "0";filename
%'lower.jpg';file-properties "XNPEU";}}

%\begin{figure}[htp]
%\centering
%\includegraphics[width=3in]{lower.jpg}
%\caption{}\label{Fig2}
%\end{figure}

%\FRAME{ftbpFU}{3.1384in}{2.6922in}{0pt%
%}{\Qcb{Range of $\Delta _{4,2,1}$ and $\Delta _{4,3,2}$ with orientation ( $%
%n=4$). The range $\Psi $ applied to these simplices give lower bounds on $%
%H_{3}.$}}{\Qlb{Fig2}}{upper.jpg}{\special{language "Scientific Word";type
%"GRAPHIC";maintain-aspect-ratio TRUE;display "USEDEF";valid_file "F";width
%3.1384in;height 2.6922in;depth 0pt;original-width 8.3662in;original-height
%7.1736in;cropleft "0";croptop "1";cropright "1";cropbottom "0";filename
%'upper.jpg';file-properties "XNPEU";}}

}

{\normalsize It is possible to describe the situation in more detail. Let }$%
\Phi$ {\normalsize denote the map}%
\begin{equation*}
P\rightarrow\left( 
\begin{array}{c}
H_{\alpha_{1}}\left( P\right) \\ 
H_{\alpha_{2}}\left( P\right)%
\end{array}
\right) .
\end{equation*}
{\normalsize Then }$\Phi$ {\normalsize restricted to }$\bigoplus
\limits_{m=3}^{n}\Delta_{m,m-1,1}$ {\normalsize is a homeomorphism. If }%
\begin{equation*}
\Phi\left( P\right) =\left( 
\begin{array}{c}
a \\ 
b%
\end{array}
\right)
\end{equation*}
{\normalsize then there exist a unique }$m$ {\normalsize and unique weights }%
$x,y,z\geq0$ {\normalsize that sum up to 1 such that }$P=x\cdot U_{m}+y\cdot
U_{m-1}+z\cdot U_{1}${\normalsize . For any distribution }$Q$ {\normalsize %
with }$\Phi\left( Q\right) =\left( 
\begin{array}{c}
a \\ 
b%
\end{array}
\right) $ {\normalsize we have }$H_{\alpha_{3}}\left( Q\right) \leq
H_{\alpha_{3}}\left( P\right) .$ {\normalsize Thus, }$\bigoplus
\limits_{m=3}^{n}\Delta_{m,m-1,1}$ {\normalsize gives tight upper bounds on }%
$H_{\alpha_{3}}$ {\normalsize in terms of }$H_{\alpha_{1}}$ {\normalsize and 
}$H_{\alpha_{2}}.$ {\normalsize We notice that this upper bound does not
depend on }$n.$ {\normalsize Similarly, the lower bound on }$H_{\alpha_{3}}$ 
{\normalsize for fixed }$H_{\alpha_{1}}$ {\normalsize and }$H_{\alpha_{2}}$ 
{\normalsize is determined by the surface }$\bigoplus
\limits_{m=2}^{n-1}\Delta_{n,m,m-1}$ {\normalsize and just as in the case of
two R\'{e}nyi entropies the lower bound will depend on }$n.$

\section{\protect\normalsize \hs DISCUSSION}

{\normalsize The result can be seen as a generalization of the result in 
\cite{Harremoes2001h}. The essential step in the whole construction is the
positivity of the generalized Vandermonde determinant. Therefore the
construction can be iterated so that one in principle can determine the
boundary of the range of any number of R\'{e}nyi entropies of positive order.%
}

\section*{\protect\normalsize ACKNOWLEDGEMENT}

{\small I thank Karol Zyczkowski for useful discussions. His paper \cite%
{Zyczkowski2003} was an important inspiration for this article. He, Flemming
Tops\o e and Christian Schaffner have also contributed with several
important remarks to this paper.\\
The author was supported by grants from
Villum Kann Rasmussen Foundation, The Banach Center, INTAS (project 00-738),
Danish Natural Research Council, and the European Pascal Network of
Excellence. }

\begin{flushright}
{\footnotesize (Received April xx, 2009.)\thinspace\ \rule{0mm}{0mm} }
\end{flushright}

{\footnotesize 
\bibliographystyle{abbrv}
\bibliography{database1}
{\small %\bigskip \\
} }

\begin{flushright}
{\footnotesize {\normalsize 
\begin{minipage}[]{124mm}
{P. Harremo\"{e}s, Centrum Wiskunde \& Informatica, Science Park 123, 1090 GB Amsterdam, Noord-Holland, The
Netherlands.
\\ E-mail: P.Harremoes@cwi.nl}
\end{minipage}
} }
\end{flushright}

\end{document}